\documentclass[12pt]{article}

\usepackage{amsmath,amssymb,amsthm}

\newcommand{\VEV}[1]{\left\langle #1 \right\rangle}

\newcommand{\diag}{{\rm diag}}

\newcommand{\GeV}{\mbox{GeV}}

\newcommand{\ie}{{\it i.e.}}
\newcommand{\eg}{{\it e.g.}}

\newcommand{\Z}[1]{{\mathbb Z}_#1}

\newcommand{\tr}{{\rm {tr}}}
\newcommand{\supP}[1]{^{(#1)}}

\newcommand{\bequ}{\begin{equation}}
\newcommand{\eequ}{\end{equation}}
\newcommand{\beqn}{\begin{eqnarray}}
\newcommand{\eeqn}{\end{eqnarray}}
\newcommand{\Ls}{\left(}
\newcommand{\Rs}{\right)}
\newcommand{\half}{{\frac12}}
\begin{document}

\begin{flushright}
MISC-2011-10
\end{flushright}

\begin{center}
{\LARGE\bf Doublet-Triplet Splitting in an $SU(5)$ Grand Unification}

\vskip 1.4cm

{\large   
Toshifumi Yamashita\footnote{e-mail: tyamashi@cc.kyoto-su.ac.jp}
}
\\

\vskip 1.0cm

{\it 
Maskawa Institute for Science and Culture, Kyoto Sangyo University, \\
Kyoto 603-8555, Japan%
}\\

\vskip 1.5cm

\begin{abstract}
We present a new solution to the doublet-triplet splitting problem which also works 
 in a supersymmetric $SU(5)$ grand unified theory and is testable 
 at TeV-scale collider experiments. 
In our model, the $SU(5)$ symmetry is broken through the Hosotani mechanism. 
Thanks to the phase nature of the Hosotani-breaking, the ``missing VEV" 
 can be realized even in an $SU(5)$ model. 
A general and distinctive prediction of this solution is the existence of light adjoint 
 chiral supermultiplets with masses of the supersymmetry breaking scale. 
Though these fields disturb the gauge coupling unification, it can be recovered keeping 
 the unified gauge coupling constant perturbative by introducing additional 
 vector-like particles, which may be also observed in the upcoming collider experiments. 
\end{abstract}

\end{center}

\vskip 1.0 cm

\newpage
%%%%%%%%%%%%%%%%%%%%%%%%%%%%%%%%%%%%%%%%%%%%%%%%%%%%%%%%%%%%%%%%
%
\section{Introduction}
%
%%%%%%%%%%%%%%%%%%%%%%%%%%%%%%%%%%%%%%%%%%%%%%%%%%%%%%%%%%%%%%%%

The grand unified theories (GUTs)~\cite{GUT} unify the three forces in the standard 
 model (SM), leading to the beautiful unification among the quarks and leptons. 
In the supersymmetric (SUSY) version~\cite{SUSY-GUT}, 
 where the so-called hierarchy problem is solved, 
 the minimal model predicts the observed gauge coupling unification (GCU) 
 in the minimal SUSY SM (MSSM), 
 up to the threshold corrections. 
Thus, the SUSY-GUTs are regarded as interesting candidates for the physics beyond 
 the SM.

They have, however, some awkward issues to address. 
The most serious one is the doublet-triplet (DT) splitting problem%
\footnote{
See Refs.~\cite{sliding,MP,flipped,DW,pNG,orbifoldGUTs,FatBrane} for known solutions.
}. 
This is caused by the failure of unifying the SM Higgs field which forces to 
 introduce its partner. 
In the minimal model, an $SU(5)$ model, its partner is color triplet to compose 
 the ${\bf5}$ representation. 
Since the triplet Higgs field induces proton decay, it has to be superheavy, while 
 the doublet Higgs field should have a mass of the electroweak scale. 
Thus, the doublet and the triplet must be split. 
In addition, the triplet mass should be much larger than the GUT scale indicated 
 by the GCU, $M_G\sim10^{16}\GeV$, 
 to sufficiently suppress the proton decay~\cite{PDrate} 
 (if we assume no tunings~\cite{PDwithFT}), 
 which would result in rather large threshold corrections.

We have recently proposed an interesting GUT scenario~\cite{gGHU}, called grand 
 gauge-Higgs unification, in which the unified gauge symmetry is broken 
 via the Hosotani mechanism~\cite{Hosotani}. 
This mechanism works in a higher-dimensional gauge theory with non-trivial cycles: 
 the extra-dimensional components of the gauge field acquire non-vanishing 
 vacuum expectation values (VEVs) to break the gauge symmetry. 
Our main point in the present paper is that the VEV discussed in Ref.~\cite{gGHU} which 
 breaks the $SU(5)$ unified group to the SM one 
 is also useful to solve the DT splitting problem. 
This is because, interestingly, the VEV has a form to be called the 
 ``missing VEV" (though in an inverse way) in the sense that it contributes 
 only to the doublet mass and not to the triplet mass. 
In fact, we will see that 
 when we introduce a bulk field in ${\bf5}$ representation with a certain boundary 
 condition (BC), its doublet component becomes massless on the background 
 while the triplet component has a mass of the compactification scale, and thus 
 the splitting is realized. 
Note that usually such a form is forbidden by the traceless condition of the adjoint 
 representation of the $SU(5)$ group. 
In contrast, in the Hosotani mechanism, the adjoint Higgs field appears as the 
 Wilson line phase and its phase nature allows the form. 
Namely, the Hosotani-breaking plays an essential role in our solution. 

Since only non-SUSY examples are discussed in Ref.~\cite{gGHU}, we examine how we 
 can realize the same VEV in SUSY models, as well as its stability. 
{}Fortunately or unfortunately, once the VEV is selected, the situation is very 
 similar to the orbifold GUT models~\cite{orbifoldGUTs} though there are several 
 constraints, \eg\ the unified symmetry is broken only by the VEV but not by brane 
 localized interactions. 
{}For instance, the mass spectra of the bulk fields are controlled by the parity of 
 the hypercharge, and the proton decay through the dimension 4 and 5 operators can be 
 suppressed by an approximate $U(1)_R$ symmetry which is consistent with the superheavy 
 color triplet Higgs field. 
There is, however, one significant difference, besides a theoretical advantage 
 that the symmetry breaking pattern in our model is well controlled by the calculable 
 dynamics independent of the ultraviolet completion~\cite{gGHU}. 
An immediately apparent by-product of the supersymmetrization is the existence of 
 light adjoint chiral multiplets with respect to the SM gauge group, 
 as the Wilson line phase gets only a loop induced mass which is suppressed by the 
 SUSY breaking scale and the mass differences relative to its superpartners 
 can not be larger than the SUSY breaking scale. 
This is a generic and characteristic prediction of our solution%
\footnote{
To be more precise, it is a prediction of SUSY grand gauge-Higgs unifications, and 
 is shared with the models with Dirac gaugino 
 masses~\cite{DiracGaugino,PolchinskiSusskind,DiracGauginoGCU}.}. 
Although these fields destroy the success of the GCU, 
 it is possible to recover it by introducing additional fields, without making the 
 unified gauge coupling constant non-perturbative below the GUT scale. 
Some of these additional fields may be also light to be observed in the TeV-scale 
 collider experiments.

This paper is organized as follows. 
In section~\ref{Sec:setup}, we briefly review the non-SUSY grand gauge-Higgs 
 unification. 
Then we supersymmetrize it to show the DT splitting can be realized without fine-tuning 
 if a certain form of the VEV of the Wilson line phase is assumed. 
We discuss the stability of the VEV in section~\ref{Sec:stability}, 
 which justifies our solution. 
In section~\ref{Sec:RelatedTopics}, we briefly comment on some related topics.
Section~\ref{Sec:summary} is devoted to the summary and discussions.

%%%%%%%%%%%%%%%%%%%%%%%%%%%%%%%%%%%%%%%%%%%%%%%%%%%%%%%%%%%%%%%%
%
\section{Setup}
\label{Sec:setup}
%
%%%%%%%%%%%%%%%%%%%%%%%%%%%%%%%%%%%%%%%%%%%%%%%%%%%%%%%%%%%%%%%%

In this section, we introduce our setup, using the simplest example of a 
 five-dimensional (5D) $SU(5)$ model compactified on an $S^1/\Z2$ orbifold 
 with its radius being of the GUT scale. 
We first review the non-SUSY version discussed in Ref.~\cite{gGHU} for illustration 
 purpose, and then supersymmetrize it.

%%%%%
\subsection{non-SUSY grand gauge-Higgs unification}
%%%%%

At first glance, there is a difficulty in the application of the Hosotani mechanism 
 to GUTs since the massless adjoint fields with respect to the remaining gauge symmetry 
 in the extra-dimensional components
 tend to be projected out in models that realize the chiral fermions. 
In Ref.~\cite{gGHU}, this difficulty is evaded by virtue of the diagonal embedding 
 method~\cite{DiagonalEmbedding}.
To realize this method in an $S^1/\Z2$ model, 
 we prepare two copies of the gauge symmetry which are exchangeable. 
Namely, we impose $SU(5)\times SU(5)\times \Z2$ symmetry for our $SU(5)$ model 
 where the $\Z2$ action exchanges the two gauge groups.
We call the gauge fields for the two $SU(5)$ groups $A\supP1_M$ and $A\supP2_M$, 
 respectively, where $M=\mu(=0\mbox{-}3),5$ is a 5D Lorentzian index.
The BCs around the two endpoints of the $S^1/\Z2$, 
 $y_0=0$ and $y_\pi=\pi R$, are given as 
\bequ
% A_M\supP1(y_i-y) = (-1)^M A_M\supP2(y_i+y),
 A_\mu\supP1(y_i-y) = A_\mu\supP2(y_i+y),\qquad
 A_5\supP1(y_i-y) = -A_5\supP2(y_i+y),
\label{BCinHosotani}
\eequ
 for $i=0,\pi$, where $y$ is the 5th dimensional coordinate. 
% and $(-1)^M=1$ ($-1$) for $M=\mu$ ($5$).
Defining the eigenstate of the exchange as $X^{(\pm)}=(X\supP1\pm X\supP2)/\sqrt2$, 
 we see that $A_\mu\supP+$ and $A_5\supP-$ obey the Neumann BC at the both endpoints 
 and thus have the zero-modes. 
In other words, the gauge symmetry of the 4D effective theory is the diagonal part of 
 $SU(5)\times SU(5)$ (or our GUT symmetry is {\it embedded} into the {\it diagonal} 
 part), and an adjoint scalar field is actually obtained. 

Since $A_5\supP-$ is a part of the gauge fields, it is not a simple adjoint scalar 
 field but composes the Wilson loop 
\bequ
W={\cal P}\exp\Ls i\int_{0}^{2L} \frac{g}{\sqrt2}{{A_5\supP-}^a(T_1^a-T_2^a)} dy\Rs
  \underrightarrow{\mbox{  on $(\bf5,\bf1)$  }}
  \exp\Ls i\diag\Ls\theta_1,\theta_2,\theta_3,\theta_4,\theta_5\Rs\Rs,
\eequ
 where ${\cal P}$ denotes the path-ordered integral, $g$ is the common gauge coupling 
 constant, $L=y_\pi-y_0=\pi R$ is the length of the extra-dimension,  
 $T_i$ is the generator of each $SU(5)$ symmetry, $a$ is an $SU(5)$ adjoint index, 
 and $\sum\theta_i=0$. 
In the last expression, we have used the (remaining) $SU(5)$ rotation to diagonalize 
 $A_5\supP-$ and employed the representation acting on the $(\bf5,\bf1)$ 
 for concreteness, where 
 $\theta_i=\sqrt2gL({A_5\supP-}^aT_{\bf5}^a)_{ii}$ 
 and $T_{\bf5}$ is the $SU(5)$ generator on ${\bf5}$ 
 with the usual normalization 
 $\tr(T_{\bf5}^aT_{\bf5}^b)=\delta^{ab}/2$%
\footnote{
The superficial difference of the factor 2 compared with the expression 
 in Ref.~\cite{gGHU} comes from the generator.
}.
As evident from the above expression, the VEV (and actually the system itself) 
 is periodic in $\theta_i$. 

The form of the VEV which is discussed in Ref.~\cite{gGHU} and we are interested in 
 is given by $\theta_1=\theta_2=\theta_3=2\pi$ and $\theta_4=\theta_5=-3\pi$, 
 \ie\ $\VEV{W}=\diag(1,1,1,-1,-1)\equiv P_W$.
This VEV does not affect the triplet component of ${\bf5}$ but does affect the doublet 
 to {\it split} it.
This ``missing VEV", which is forbidden for a simple adjoint scalar field by the 
 traceless condition, is allowed thanks to the phase nature of the Hosotani-breaking.

This system with the non-trivial VEV of the Wilson line is known to be equivalent 
 to the one connected by the (broken) gauge transformation, especially with the gauge 
 parameter $\alpha\supP-=g\VEV{A_5\supP-}(y-y_0)$
 by which the VEV is gauged away~\cite{HHHK}. 
The former is called as the Hosotani basis and the latter as the Scherk-Schwartz (SS) 
 basis where the effect of the VEV appears in the BCs around $y=y_\pi$, modified 
 from Eq.~\eqref{BCinHosotani} to
\bequ
% A_M\supP1(y_\pi-y) = (-1)^MP_WA_M\supP2(y_\pi+y)P_W^\dagger.
 A_\mu\supP1(y_\pi-y) = P_WA_\mu\supP2(y_\pi+y)P_W^\dagger, \qquad
 A_5\supP1(y_\pi-y) = -P_WA_5\supP2(y_\pi+y)P_W^\dagger.
\label{BCinSS}
\eequ
With these modified BCs, the $SU(5)$ symmetry is broken down to the SM one, and 
 the SM adjoint components of $A_5$ have the zero-modes. 
The components corresponding to the broken generators do not have the zero-modes 
 in this basis, while they are eaten through the Higgs mechanism in the Hosotani basis. 
Note that the way of the symmetry breaking in the SS picture, \ie\ by the BCs, 
 is the same as in the orbifold breaking~\cite{orbifoldGUTs}. 
Actually the situation is quite similar to the orbifold GUTs although 
 with several constraints, \eg\ on the possible matter content.

{}For the matter fields, in this paper, we treat only those that are non-singlet 
 of at most one of the gauge group, for simplicity. 
To be more concrete, we introduce for instance a fermion 
 $\Psi({\bf R},{\bf1})\supP1$ with ${\bf R}$ being a representation of the $SU(5)$ 
 group%
\footnote{
{}For the other case where $\Psi({\bf R}_1,{\bf R}_2)$ is introduced, 
 see Ref.~\cite{gGHU}.
}. 
Then, the exchange $\Z2$ symmetry requires its $\Z2$ partner 
 $\Psi({\bf 1},{\bf R})\supP2$ as well.
Their BCs are given as 
\bequ
 \Psi\supP1(y_\pi-y) = -\eta_i^{\Psi}\gamma_5\Psi\supP2(y_\pi+y),
\label{FermionBCinHosotani}
\eequ
 in the Hosotani basis where $\eta_i=\pm1$ is a parameter associated with each fermion. 
As one of $\eta_i$ can be reabsorbed by changing $\gamma_5$, \ie\ by the charge 
 conjugation, we set $\eta_0=+1$ and $\eta_\pi=\eta$ hereafter. 
Then, $\Psi_L\supP+$ and $\Psi_R\supP-$ have the zero-modes when $\eta=+1$ while 
 none have when $\eta=-1$. 
Thus, the zero-modes appear always in vector-like pairs of $\Z2$ even and odd fields 
 from the bulk fermion, and we put chiral fermions on a brane. 

In the SS basis, the BCs become 
\bequ
 \Psi\supP1(y_\pi-y) = -\eta_i^{\Psi}\gamma_5 W_{\bf R}\Psi\supP2(y_\pi+y),
\label{FermionBCinSS}
\eequ
 where $W_{\bf R}$ is the Wilson line phase acting on ${\bf R}$.
It is easy to derive $\Psi\supP1(y+2L)=\eta W_{\bf R}\Psi\supP1(y)$, and 
 we call the components with $\eta W_{\bf R}=1$ ($-1$) (anti-)periodic.

In particular, for ${\bf R}={\bf 5}$ with $\eta=-1$, the doublet component has 
 the zero-mode while the triplet does not. 
Although we can get a massless doublet Higgs scalar field at the tree level 
 in a similar way, 
 the loop corrections likely make it superheavy, and thus we consider SUSY models.

%%%%%
\subsection{SUSY version}
%%%%%

The same story discussed in the previous subsection can be applied also in SUSY models
 if we replace all the fields by the corresponding superfields. 
Thus, once the desired VEV $P_W$ is obtained, the DT splitting is easily realized 
 by introducing a bulk ${\bf 5}$ hypermultiplet with $\eta=-1$ for the Higgs fields. 

Then, the remaining task is to examine when the VEV is realized. 
Although, according to the literature~\cite{NoGlobal}, 
 it is difficult to realize the vacuum as the global minimum, 
 the vacua on a local minima bring no problems if their lifetimes are long enough. 
Actually, once the universe cools down on a local minimum, in our case, 
 since the length of the ``tunnel" is of order $1/R\sim M_G$,
 the tunneling rate is parametrically suppressed. 
Thus, we examine only that the vacuum lives on a minimum and do not care if it is 
 the global one or not in the following. 
For this purpose, we should check if there are no tadpole terms for the fluctuations 
 of $\theta_i$ around the desired vacuum, $\delta\theta_i$, and 
 if they are not tachyonic. 

Note that $\delta\theta_i$ is odd under the exchange $\Z2$ symmetry by the same reason 
 as the so-called H-parity~\cite{H-parity}: 
 since the system is invariant under $\theta_i\to-\theta_i$ and 
 $\theta_i\to\theta_i+2\pi$, so is under $\delta\theta_i\to-\delta\theta_i$ 
 even for $i=4,5$ for which $\VEV{\theta_i}$ is non-trivial as 
 $\theta_i=\pi+\delta\theta_i\sim-(\pi+\delta\theta_i)\sim\pi-\delta\theta_i$% 
\footnote{
It is also understood by the transformation of the Wilson line under the $\Z2$ action 
 $W\to W^*$ and the fact that the VEV $\VEV{W}$ is real and thus invariant under it.
}. 
This $\Z2$ invariance protects the tadpole terms even though there is a SM singlet 
 chiral multiplet as the adjoint of the $U(1)$ hypercharge which couples 
 both to heavy and light fields, in great contrast to the sliding singlet 
 mechanism applied to $SU(5)$ models~\cite{PolchinskiSusskind,tadpole}%
\footnote{
The other ``missing VEV" which affects only the triplet mass, \ie\ 
 $\theta_1=\theta_2=\theta_3=4\pi/3$ and $\theta_4=\theta_5=-2\pi$, 
 breaks the $\Z2$ symmetry and thus huge tadpole term will be induced by the quantum 
 corrections. 
}.
Then, the question is only the signs of the mass squared of $\delta\theta_i$.

%%%%%%%%%%%%%%%%%%%%%%%%%%%%%%%%%%%%%%%%%%%%%%%%%%%%%%%%%%%%%%%%
%
\section{Stability of the VEV}
\label{Sec:stability}
%
%%%%%%%%%%%%%%%%%%%%%%%%%%%%%%%%%%%%%%%%%%%%%%%%%%%%%%%%%%%%%%%%

In this section, we study the mass squared of $\delta\theta_i$ to find if the 
 vacuum is stable or unstable. 
Since the mass terms are generated only by the loop effects which are vanishing 
 if the SUSY is exact, the signs crucially depend on the SUSY breaking. 
First, as a simple example, we employ the SS SUSY breaking~\cite{ScherkSchwartz}, 
 and then examine the condition for the vacuum to be stable, 
 with general SUSY breaking.

In the SS mechanism, the SUSY is broken by BCs which twist the $R$-symmetry and thus 
 give different masses for different components of a supermultiplet.
By this, the fermion (scalar) component in each gauge (hyper-) multiplet gets 
 SUSY breaking masses $\beta/R$ with $\beta$ being a twist parameter, 
 called the SS phase. 
Although a tiny $\beta$ seems unnatural since, in the supergravity, this breaking is 
 equivalent to the one by the radion $F$-term~\cite{SSPhase}, 
 we assume $\beta\sim10^{-14}$ by hand 
 as we use this mechanism just as an illustrating example. 

With this SUSY breaking, the contributions from each periodic and anti-periodic modes 
 to the 1-loop effective potential of the fluctuations $\delta\theta_i$ are given 
 by~\cite{1-loopPot-SS} 
\beqn
&& V\supP+(\delta\theta_i)=
 2sC\sum_{w=1}^\infty\frac1{w^5}(1-\cos(2\pi w\beta))\cos(wQ_i\delta\theta_i)
 \sim2sC\half\Ls\ln(4\beta^2)-3\Rs \beta^2(Q_i\delta\theta_i)^2,\quad
\\
&& V\supP-(\delta\theta_i)=
 2sC\sum_{w=1}^\infty\frac{(-1)^w}{w^5}(1-\cos(2\pi w\beta))\cos(wQ_i\delta\theta_i)
 \sim2sC(\ln2)\beta^2(Q_i\delta\theta_i)^2,
\eeqn
 respectively, where $s=-1$ ($+1$) for the gauge (hyper-) multiplets, 
 $C=3/(64\pi^7R^5)$, $Q_i$ is the charge of the relevant mode with respect to the 
 $U(1)$ symmetry corresponding to $\delta\theta_i$, 
 and we have expanded the functions for $\beta\ll1$~\cite{LEET} 
 at the last expressions.
When $\beta\ll1$, the contributions to the mass of $\delta\theta_i$ are dominated by 
 those that are enhanced by the factor $\ln(4\beta^2)$ from the periodic modes, 
 to be more precise, from the zero-modes. 
Given that $\beta$ is so small, the higher-loop corrections are not so suppressed
 and we should resum all the leading log terms, by solving the renormalization group 
 equations (RGEs). 
The other terms are treated as the threshold corrections which are of the next to 
 the leading order. 
Thus, if we are satisfied with the 1-loop RGE approximation, the mass of $\delta\theta_i$ 
 is controlled only by the 4D effective theory~\cite{LEET}. 
This observation is apparently not specific to the SS breaking but is applied to 
 the cases with general SUSY breaking.

We shall turn to the RGEs in models with the adjoint chiral 
 multiplets, $(\Sigma_8,\,\Sigma_3,\,\Sigma_1)$ for $(SU(3),\,SU(2),\,U(1))$ of the 
 SM and possible vector-like pairs for the GCU 
 in addition to the MSSM matter content. 
The adjoint multiplets interact with the bulk fields via the Yukawa terms in the 
 superpotential given in Ref.~\cite{higherSUSY}. 
The RGEs can be calculated, for example as in Ref.~\cite{nrMSSU(5)}, 
 and we see that
 the soft mass squared terms of $\Sigma_8$ and $\Sigma_3$ 
 are pushed up by the 
 gaugino contributions, and thus they are likely positive. 
On the other hand, $\Sigma_1$ is gauge singlet and 
 its soft mass squared 
 is driven only by the Yukawa interactions. 
Then, to make the contribution positive, some of the soft mass squared 
 of the superfields that take part in the Yukawa interactions have to be negative. 
It might sound unnatural, but it says just that the fermion component is heavier than 
 the scalar one, and the latter needs not tachyonic. 
In fact, for instance, while $\delta\theta_i$ is massless due to the shift symmetry 
 at the tree level, 
 its fermion partner may have 
 a mass of the SUSY breaking scale. 
This situation is expressed by a $\mu$-like term of the SUSY breaking scale 
 and a soft mass squared term that cancels the contribution of the $\mu$-like term 
 to the $\delta\theta_i$ mass by a negative coefficient, 
 in the superfield formalism~\cite{NegativeM2inSS}. 
If the supersymmetric mass is larger, as the cases of the supermultiplets 
 with masses of the intermediate scale which may be introduced to recover the GCU, 
 the scalar component is also massive. 
{}For these multiplets, the magnitudes of the soft mass squared may be rather large 
 and then the soft masses of $\delta\theta_i$ become large enough to avoid the 
 tachyonic masses against possible $B$-terms.

In this way, by choosing appropriate SUSY breaking pattern, which does not require 
 fine-tuning, it is possible that all the $\delta\theta_i$ have positive mass squared 
 to make the vacuum stable. 
As discussed, once the vacuum is realized, its lifetime is parametrically long 
 even if it resides on a local minimum. 
Then our analysis on the vacuum which leads to the DT splitting even in an $SU(5)$ model 
 is valid.

%%%%%%%%%%%%%%%%%%%%%%%%%%%%%%%%%%%%%%%%%%%%%%%%%%%%%%%%%%%%%%%%
%
\section{Some related topics}
\label{Sec:RelatedTopics}
%
%%%%%%%%%%%%%%%%%%%%%%%%%%%%%%%%%%%%%%%%%%%%%%%%%%%%%%%%%%%%%%%%

Before closing this paper, we shall comment on the Yukawa couplings, the proton decay, 
 the GCU and the $\mu$-problem. 
Many of them have a similar feature to the orbifold GUTs~\cite{orbifoldGUTs}. 

As mentioned above, in our setup, chiral fermions can not be obtained from the 
 bulk fields, and thus we put the three generational quarks and leptons on a brane, 
 where the $SU(5)$ symmetry is not broken. 
To evade the wrong GUT relation between the down-type quarks and the charged leptons, 
 we have to introduce bulk matter fields as messengers of the $SU(5)$ 
 breaking~\cite{orbifoldGUTs}. 
{}For instance, a bulk $\bf{\bar5}$ hypermultiplet with $\eta=-1$ serves an additional 
 lepton doublet and its vector-like partner as the zero-modes. 
Since the former (latter) is even (odd) under the $Z_2$ symmetry that forbids the tadpole 
 term of $\Sigma_1$, they can not be superheavy, while they can be if we introduce 
 a further bulk $\bf{5}$ hypermultiplet. 
In any case, these additional matters can be used to break the wrong relation because 
 some of the zero-modes can couple to the boundary lepton doublets, 
 while there are no additional light right-handed down-type quarks. 
Similarly, a pair of bulk ${\bf10}$ and $\bf{\bar{10}}$ could be used.

At this stage, we do not attempt to explain the hierarchical structure of the Yukawa 
 couplings, but just mention that the flavor 
 symmetries~\cite{anomalousU(1), discreteFS} do not conflict with our solution. 
We leave the task to construct a realistic model of the flavor as a future work.

The proton decay via the dimension 4 and 5 operators is efficiently suppressed 
 by an approximate $U(1)_R$ symmetry. 
The one via the gauge boson exchange is, on one hand, enhanced because all the 
 Kaluza-Klein (KK) modes contribute and their couplings to the boundary fields are 
 larger by the factor $\sqrt2$ than those of the corresponding zero-modes 
 due to the normalizations of the 5th dimensional wave functions.
On the other hand, if the bulk originated field discussed above dominates a light 
 mode while its $SU(5)$ partner connected by the gauge boson decouples due to 
 the KK mass, it is suppressed. 
Thus, it highly depends on models of the flavor and, unfortunately, it seems 
 difficult to derive a generic prediction. 
These features are shared with the orbifold GUTs~\cite{orbifoldGUTs}.

The GCU in the MSSM is neither a prediction in our scenario, 
 since the light adjoint chiral multiplets $\Sigma_1$, $\Sigma_3$ and $\Sigma_8$ exist, 
 which is characteristic of our scenario. 
They contribute to the beta function coefficient by $(3,2,0)$ 
 for the $(SU(3),\,SU(2),\,U(1))$
 with the $SU(5)$ normalization of the $U(1)$. 
To recover the unification, we shall add $(n,1+n,3+n)$ in some way%
\footnote{
It is possible that the unification scale is also modified~\cite{DiracGauginoGCU}.
}. 
The straightforward possibility is to introduce their $SU(5)$ partners, 
 $({\bf3},\,{\bf2},\,5/6)$ and $({\bf{\bar3}},\,{\bf2},\,-5/6)$ 
 which leads to $n=2$. 
In this case, however, a Landau pole appears below the GUT scale. 
An example with $n=1$ which keeps the couplings perturbative, 
 $\alpha_{\rm GUT}\sim0.3$ at the 1-loop level, is to introduce 
 two pairs of $({\bf1},\,{\bf2},\,1/2)$ and each pair of $({\bf{\bar3}},\,{\bf1},\,-2/3)$ 
 and $({\bf1},\,{\bf1},\,1)$.
These pairs can originate from bulk ${\bf5}$ multiplets with $\eta=-1$ and 
 ${\bf10}$ with $\eta=+1$.
Some of these may be identified with the bulk fields needed to break the wrong GUT 
 relation or the messengers of the SUSY breaking.

Finally, we shall discuss the $\mu$-problem. 
The $U(1)_R$ symmetry that suppresses the proton decay forbids the $\mu$-term, 
 which could explain why the $\mu$-term is of the SUSY breaking 
 scale~\cite{orbifoldGUTs}. 
In addition, however, when we regard the additional two pairs of the doublet 
 as the matter or messenger fields, the two Higgs doublets should come from one 
 bulk field and 
 have the different $\Z2$ parities which also forbid the $\mu$-term in our scenario. 
Then, since the $\Z2$ symmetry should not be broken so strongly to protect the 
 tadpole term, the $\mu$-term may be too suppressed.
We may consider that proton decay is suppressed not by the $U(1)_R$ but by the $\Z2$ 
 symmetry to avoid it. 

We can also consider another case where the $\Z2$ symmetry remains unbroken. 
In this case, one pair of the additional doublet should be regarded as Higgs fields 
 in order to write the $\mu$-term (after the $U(1)_R$ breaking). 
To write the mass terms for the other additional pairs, we should double them and 
 set their masses to be of the intermediate scale to effectively reproduce 
 the required contributions to the GCU. 
In this case, the model is a SUSY version of the inert 
 singlet~\cite{ISM}/doublet~\cite{IDM}/triplet~\cite{ITM} 
 coexisting model, with the color octet chiral multiplet in addition, 
 which can decay 
 only through virtual superheavy fields and thus may form the so-called R-hadron.

%%%%%%%%%%%%%%%%%%%%%%%%%%%%%%%%%%%%%%%%%%%%%%%%%%%%%%%%%%%%%%%%
%
\section{Summary and Discussions}
\label{Sec:summary}
%
%%%%%%%%%%%%%%%%%%%%%%%%%%%%%%%%%%%%%%%%%%%%%%%%%%%%%%%%%%%%%%%%

We have presented a new solution to the DT splitting problem in SUSY-GUTs where 
 the unified gauge symmetry is broken through the Hosotani mechanism~\cite{Hosotani}. 
An adjoint chiral multiplet is obtained by the diagonal embedding 
 method~\cite{DiagonalEmbedding,gGHU} on an orbifold. 
Thanks to the fact that it is not an simple adjoint field but composes the Wilson line 
 phase, the ``missing VEV" is allowed even in $SU(5)$ models, which is a key ingredient 
 of our solution. 
We have discussed that even if the vacuum resides on a local minimum, 
 the lifetime is parametrically long and then we examined the stability of the vacuum. 
We have showed that once the VEV is realized, fortunately or unfortunately, our scenario 
 basically reduces to the orbifold GUTs~\cite{orbifoldGUTs} with several constraints. 
Thus, some discussions, \eg\ on the Yukawa couplings and the proton decay, 
 are shared in these scenarios. 

A distinctive difference appears in the possible matter content in the 4D effective 
 theories. 
In particular, our scenario predicts light adjoint chiral multiplets with respect to 
 the SM gauge group which are testable at TeV-scale collider experiments. 
Although these light fields disturb the GCU, it can be recovered by introducing 
 additional vector-like fields, 
 keeping the unified gauge coupling constant perturbative. 
Depending on the masses of the additional colored particles, the lifetime of the 
 color octet multiplet may become rather long to form the so-called R-hadron. 

In addition, our models contain a $\Z2$ symmetry under which the light adjoint multiplets 
 change the sign. 
The $\Z2$ symmetry may be broken slightly and may remain exact. 
The phenomenology of the colorless fields would largely depends on the presence of 
 the breaking. 
{}For example, when the $\Z2$ symmetry remains unbroken, in order to write the 
 $\mu$-term, two inert doublet fields~\cite{IDM} should appear, 
 in addition to the inert singlet~\cite{ISM} and triplet~\cite{ITM}. 
We will study their phenomenology in another place~\cite{gGHUpheno}.

Finally, we comment on the application to the string theory. 
The idea of the diagonal embedding~\cite{DiagonalEmbedding} was proposed in the 
 context of the string theory in the first place. 
Thus, our field theoretical setup may be regarded as an effective theory of 
 such string models. 
Although, in this view, we neglect all the stringy effects which could modify our 
 discussion especially when the compactification scale is not smaller than the 
 string scale, it may be possible that our solution can be applied also to such 
 string models.

%%%%%%%%%%%%%%%%%%%%%%%%%%%%%%%%%%%%%%%%%%%%%%%%%%%%%%%%
\section*{Acknowledgments}
%%%%%%%%%%%%%%%%%%%%%%%%%%%%%%%%%%%%%%%%%%%%%%%%%%%%%%%%

We appreciate M.~Kakizaki, S.~Kenemura, K.Kojima, Y.~Nakai and K.~Takenaga
 for valuable discussions.

\end{document}